\documentclass[preprint,aps]{revtex4-2}
\usepackage{graphicx,epsfig,subfig,dcolumn,bm,mathrsfs,amsmath,amsthm,color}
\usepackage[colorlinks]{hyperref}
\usepackage{rotating}
\usepackage{CJKutf8}
\newcommand*{\Cn}[1]{\begin{CJK}{UTF8}{gbsn}#1\end{CJK}}
\newcommand*{\Ja}[1]{\begin{CJK}{UTF8}{ipxm}#1\end{CJK}}

\begin{document}
\title{Capillary Filling Dynamics in Polygonal Tubes}

\author{Chen Zhao (\Cn{赵晨})}
\affiliation{Wenzhou Key Laboratory of Biomaterials and Engineering, Wenzhou Institute, University of Chinese Academy of Sciences, Wenzhou 325000, China}
\affiliation{Institute of Theoretical Physics, Chinese Academy of Sciences,
Beijing 100190, China}

\author{Yu Huang (\Cn{黄瑜})}
\affiliation{South China Advanced Institute for Soft Matter Science and Technology, School of Emergent Soft Matter, South China University of Technology, Guangzhou 510640, China}

\author{Tingxuan Chen (\Cn{陈庭轩})}
\affiliation{South China Advanced Institute for Soft Matter Science and Technology, School of Emergent Soft Matter, South China University of Technology, Guangzhou 510640, China}

\author{Jiaxuan Li (\Cn{李嘉轩})}
\affiliation{South China Advanced Institute for Soft Matter Science and Technology, School of Emergent Soft Matter, South China University of Technology, Guangzhou 510640, China}

\author{Jiajia Zhou (\Cn{周嘉嘉})}
\email[]{zhouj2@scut.edu.cn}
\affiliation{South China Advanced Institute for Soft Matter Science and Technology, School of Emergent Soft Matter, South China University of Technology, Guangzhou 510640, China}
\affiliation{Guangdong Provincial Key Laboratory of Functional and Intelligent Hybrid Materials and Devices, South China University of Technology, Guangzhou 510640, China}

\author{Masao Doi (\Ja{土井正男})}
%\email[]{doi.masao.y3@a.mail.nagoya-u.ac.jp}
\affiliation{Wenzhou Key Laboratory of Biomaterials and Engineering, Wenzhou Institute, University of Chinese Academy of Sciences, Wenzhou 325000, China}
\affiliation{Oujiang Laboratory (Zhejiang Lab for Regenerative Medicine, Vision and Brain Health), Wenzhou 325000, China}

\date{\today}

\begin{abstract}

We study the dynamics of capillary filling in tubes of regular polygon cross-section. 
Using Onsager variational principle, we derive a coupled ordinary differential equation and partial differential equation, which respectively describe time evolution of the bulk flow and the saturation profile of the finger flow. 
We obtain both numerical solution and self-similar solution to the coupled equations, and the results indicate that the bulk flow and the finger flow both follow the $t^{1/2}$ time-scaling. 
We show that due to the coupling effect of the finger flow, the prefactor for the bulk flow is smaller than that of the Lucas-Washburn prediction. 
The reduction effect is more pronounced when the side number $n$ of the regular-polygon is small, while as $n$ increases, the prefactor approaches Lucas-Washburn prediction.

\end{abstract}

\maketitle

\section{Introduction}

Liquid transportation in narrow confined geometries is ubiquitous in nature. 
Examples are the hierarchical structural features on the peristome surface of the Nepenthes alata pitcher \cite{ChenHuawei2016} and the parallel array of xylems in a block of wood \cite{Crawford1983}, both demonstrating directional water transport. 
Capillarity becomes dominant over gravity when the length scale of the system is smaller than the capillary length.
The capillary process is a spontaneous phenomenon requiring no energy input \cite{Patrascu2022}, which naturally evolves to a state of total energy minimum.
Capillarity is wildly investigated for its crucial applications in many aspects, such as in microfluidic devices \cite{Grunze1999, Gau1999, Lai2010}, lithography \cite{Unger2000}, and biomimetic preparation \cite{Parker2001, Wang2014, Hon2008}.

The theory of capillary imbibition was pioneered by Lucas \cite{Lucas1918} and Washburn \cite{Washburn1921} one century ago. 
Their theory showed that the filling length $h(t)$ in a circular tube follows the relation
\begin{equation}
\label{LW}
  h(t) = \sqrt{\frac{\gamma a \cos \theta}{2 \eta}} t^{1/2} ,
\end{equation}
where $\theta$ is the equilibrium contact angle between the liquid and the tube's inner surface, $a$ is the radius of the tube, $\gamma$ and $\eta$ are the surface tension and the viscosity of the fluid, respectively. 
The $t^{1/2}$ scaling law is very robust, which has been verified in both micro- and nano-scale systems \cite{Dimitrov2007, Schebarchov2008, YaoYang2017, YaoYang2018}.
The prefactor in Eq.~(\ref{LW}) (sometime called Lucas-Washburn factor $C_{\mathrm{LW}}$), on the other hand, depends on the specific systems \cite{Chauvet2012, ZhaoChen2021, YaoYang2018a}. 
Lucas-Washburn theory focused on the bulk part of the liquid, as there exists one spherical meniscus whose length remains finite during the imbibition process in the tube of circular cross-section.
If the tube has sharp corners, the corner flow (we shall call it finger flow) apparently appears in experiments \cite{Ponomarenko2011, Wijnhorst2020, Kolliopoulos2021, McCraney2021, Kubochkin2022, Kim2020}, involving complex boundaries and flow fields. 
Many theoretical researches have been performed for the capillary filling of the finger flow, such as in triangular grooves \cite{Ayyaswamy1974}, the square tube \cite{Dong1995, YuTian2018, Prat2007, Chauvet2009} and the rectangular tube \cite{YuTian2021, ZhaoChen2021}. 
In recent years, the simulation studies combining both the bulk flow and the finger flow had been conducted to analyze and predict the static equilibrium state and also the dynamics \cite{Kialashaki2022, Panter2023, ThammannaGurumurthy2018}.

The Lucas-Washburn law is well established in the macroscopic scale, where the length of the system is greater than micrometer. 
With the advance in microfabrication techniques, studies started to probe nanoscale cornered systems, which showed that the $t^{1/2}$ scaling remains valid but the prefactors vary significantly. 
Most experiments observed a reduction of the prefactor with respect to the Lucas-Washburn prediction and various interpretations have been made, including the dynamic contact angle, precursor film, trapped gas bubbles, etc. \cite{Tas2004, Delft2007, Persson2007, Haneveld2008, Hamblin2011, Chauvet2012}
Recent advancements in computational methods, including machine learning techniques, have further enhanced the modeling of capillary effects in complex systems, enabling efficient predictions of phase behaviors and fluid dynamics under nanoscale confinement \cite{Zhang2020}. 
Since the finger flow appears in cornered geometrical tubes, non-negligible mutual interactions exist among the bulk and the finger flows. 
It has been demonstrated in our previous work \cite{YuTian2018, ZhaoChen2021} that the corner flows slow down the filling speed of the bulk liquid.
Here, we extend our study to the cases of regular polygon. 

The condition that the liquid wets the solid surface is $\gamma_{SV} > \gamma_{SL}$, with $\gamma_\mathrm{SV}$ and $\gamma_\mathrm{SL}$ representing the solid-vapor and solid-liquid interfacial tensions, respectively. From Young's equation, $ \gamma _{SV} - \gamma _{SL} = \gamma \cos \theta$, this condition gives $\cos\theta>0$, which corresponds to $\theta<\pi/2$. 
For corner geometries, the condition is different due to the presence of the finger flow.  
Concus and Finn \cite{Concus1969} firstly proved that for the open corner geometry of inner angle $2 \alpha$, the condition $ \alpha + \theta < \pi/2 $ must be satisfied for the finger flow to happen. 
For the capillary tube geometry that the cross-section is closed, we show the condition for finger flows remains the same. 
%We gives a plausible analysis based on our current theory.

In the present paper, we develop a theoretical model for capillary filling dynamics in tubes with regular polygonal cross-section. 
Our method is based on the Onsager variational principle \cite{DoiSoft, Doi2021}. 
We derive coupled equations, one partial differential equation (PDE) for the finger flow and one ordinary differential equation (ODE) for the bulk flow. 
The coupled equations can be solved numerically, and also can be cast into a self-similar form. 
We show both the bulk and the finger lengths follow a time scaling of $t^{1/2}$. 
Due to the presence of the finger part, the filling velocity of the bulk part is smaller than the Lucas-Washburn predication. 
This reduction effect is correlated to the starting saturation $s^*$ (a relative stable saturation of finger region when coupled the bulk part, which is detailedly discussed in Section II, part $E$) of the finger. 
As the number of the polygon sides increases, the value of $s^*$ decreases, and the reduction effect caused by the finger part is reduced.
In the limit of large number of sides, the regular-polygon becomes a circle, and the bulk speed approaches the Lucas-Washburn factor.
%\begin{figure}[htbp]
%  \centering
%  \includegraphics[width=1\columnwidth]{figs/sketch1}
%  \caption{Sketch of the cross section of regular-polygon tubes: regular triangle, square, regular pentagon, regular hexagon, regular heptagon and etc.}
%  \label{sketch1}
%\end{figure}

%%%%%%%%%%%%%%%%%%%%%%%%%%%%%%%%%%%%%%%%%%%%%%%%%%%%%%%%%%%%%%%%
\section{Theoretical analysis}

\subsection{Model description}

\begin{figure}[htbp]
  \centering
  \includegraphics[width=0.6\columnwidth]{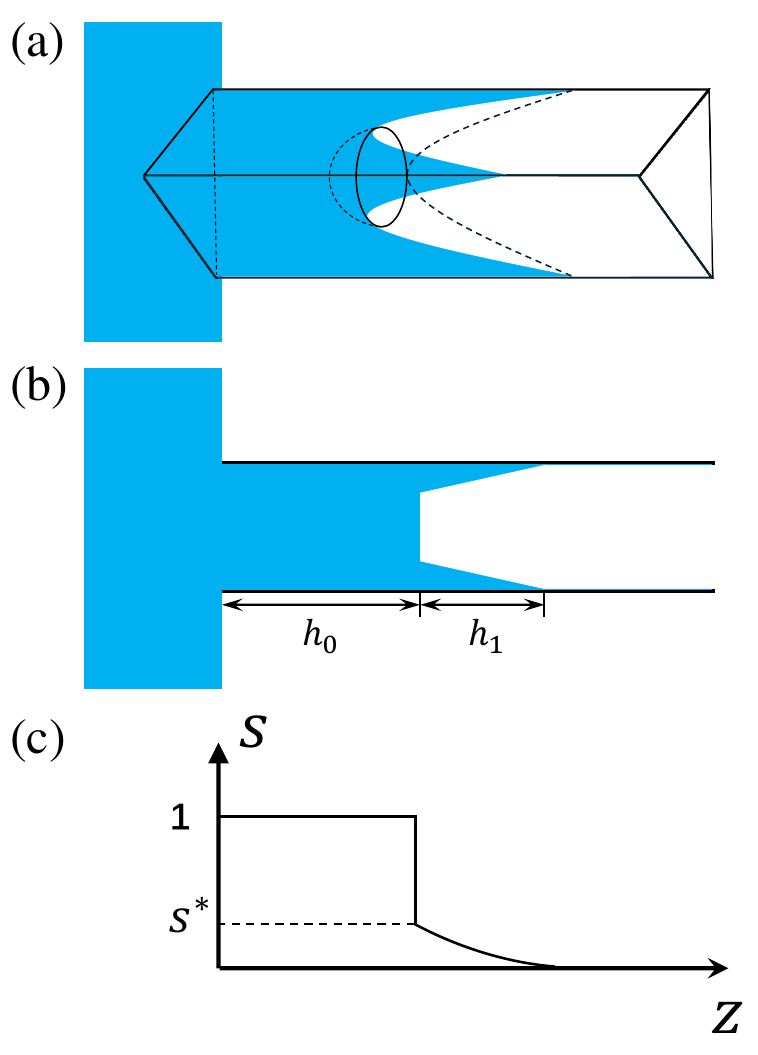}
  \caption{(a) Sketch of the capillary imbibition in the triangular tube. The liquid inside the tube can be separated into three regions: bulk part with saturation $s=1$, finger part starting with $s^*$, and the transition region in between. Our model neglects the transition region and uses the simplified model shown in (b). The saturation $s$ changes from 1 to $s^*$ discontinuously at height $h_0$. The length of the bulk and the finger are denoted by $h_0$ and $h_1$, respectively. The saturation profile $s(z)$ is also shown in (c).}
  \label{sketch1}
\end{figure}

We consider a tube with a regular polygonal cross-section that is in contact with a fluid reservoir.
The tube is placed in the horizontal direction.
We shall focus on the cases where the typical side length is less than the capillary length, thus gravity effect can be neglected.
%The inner surface of the tube is lyophilic thus the capillary force drives the fluid to imbibe into the tube.
We take $z$-axis along the tube direction, and define the saturation $s(z)$ as the ratio of the area occupied by the fluid to the total area.
A schematic picture of triangular system is illustrated in Fig.~\ref{sketch1}.
The length of the region where $s=1$ is denoted as $h_0$ (which we will call \emph{bulk}), while the length of partially-filled region ($s<1$) is $h_1$ (which we will call \emph{finger}).

Since our study focused on the long-time scaling laws as $h_0(t) \& h_1(t) \gg a$, we used an approximation where the transition region from the bulk to finger is neglected \cite{Weislogel2011,Weislogel2012}. 
We shall consider a simplified model that the partially filled region starts with a critical saturation $s^*$ [see Fig.~\ref{sketch1}(b)], which characterizes the coexistence of the fully saturated and partially saturated regions \cite{YuTian2018, YuTian2021}.
In real systems, the transition region from $s = 1$ to $s^*$ has a finite size, which is of the order of the tube's side length.
At later time in the imbibition, $h_0$ and $h_1$ become much longer than the transition region, thus we can treat the saturation to have a discontinuous jump from 1 to $s^*$ at the place where the bulk and finger meet. 

\subsection{Time evolution equations}

Onsager variational principle is a general method to derive the time-dependent equations for soft matter system without inertial effect \cite{DoiSoft}. 
The starting point is to write down a quantity called Rayleighian, that is the summation of the time derivative of the free energy and the dissipation function. 
We first start with a general model without specifying the cross-section shape, and we will implement the regular-polygon in the next subsections. 

The free energy of the system is written in the summation of the bulk contribution and the finger
\begin{equation}
\label{eq1}
  A = f(1) h_{\rm 0} + \int_{h_{\rm 0}}^{h_{\rm 0}+h_{\rm 1}} dz f(s) \, ,
\end{equation}
where $f(1)$ represent the interfacial free energy density (per unit length) of the bulk, and $f(s)$ for the finger. 
The time derivative of the free energy is 
\begin{equation}
\label{eq2}
\begin{aligned}
  \dot{A} = f(1) \dot{h}_{\rm 0} + \int_{h_{\rm 0}}^{h_{\rm 0}+h_{\rm 1}} dz f'(s) \frac{\partial s}{\partial t} - f(s^*) \dot{h}_{\rm 0} \, ,
\end{aligned}
\end{equation}
where $f'(s) \equiv \mathrm{d} f(s) / \mathrm{d} s$. The term on the tip position of the finger ($z=h_0+h_1$) vanishes due to $f(0)=0$. 

The volume conservation equation in the finger is
\begin{equation}
\label{eq3}
  \frac{\partial s}{\partial t} = - \frac{\partial {j_1(s)}}{\partial z} \, ,
\end{equation}
where $j_1(s)$ is the volume flux divided by the cross-section area in the finger.
Using the conservation equation and integration-by-part, Eq.~(\ref{eq2}) becomes
\begin{equation}
\label{eq4}
  \dot{A} = f(1) \dot{h}_{\rm 0} + f'(s^*) j_1^* + \int_{h_{\rm 0}}^{h_{\rm 0}+h_{\rm 1}} dz f''(s) \frac{\partial s}{\partial z} j_1(s) - f(s^*) \dot{h}_{\rm 0} \, ,
\end{equation}
where $f''(s) \equiv \mathrm{d}^2 f(s) / \mathrm{d} s^2$, and $j_1^* = j_1(s^*)$ is the volume flux at the entrance of the finger.

The dissipation function is half of the energy dissipated in the system. 
In general, the dissipation function has a quadratic form of the velocity, with a frictional coefficient that depends on the local saturation. 
For the bulk, the dissipation function has the form of $\frac{1}{2} \zeta(1) h_0 j_0^2$, where $\zeta(1)$ is the friction coefficient for full saturation and $j_0 = j_1^* + (1-s^*)\dot{h}_{\rm 0}$ is the flux in the bulk. 
For the finger, the flux varies for the position, and the dissipation function is written in an integral $\displaystyle \frac{1}{2} \int_{h_{\rm 0}}^{h_{\rm 0}+h_{\rm 1}} dz \zeta(s) j_1^2(s)$. 
The total dissipation function is 
\begin{equation}
\label{eq5}
  \Phi = \frac{1}{2} \zeta(1) h_0 j_0^2 + \frac{1}{2} \int_{h_{\rm 0}}^{h_{\rm 0}+h_{\rm 1}} dz \zeta(s) j_1^2(s) \, ,
\end{equation}
where $\zeta(s)$ is the friction coefficient in the finger, and it's a function of the local saturation.

The Rayleighian is then given by $\mathscr{R} = \dot{A} + \Phi$
\begin{equation}
\label{eq7}
  \mathscr{R} =  f(1) \dot{h}_{\rm 0} + \int_{h_{\rm 0}}^{h_{\rm 0}+h_{\rm 1}} dz f'(s) \frac{\partial s}{\partial t} - f(s^*) \dot{h}_{\rm 0} + \frac{1}{2} \zeta(1) h_0 j_0^2 + \frac{1}{2} \int_{h_{\rm 0}}^{h_{\rm 0}+h_{\rm 1}} dz \zeta(s) j_1^2(s) \, .
\end{equation}
The Rayleighian is written as a function of the bulk length $h_0(t)$, the finger saturation profile $s(z;t)$ for $h_0 \le z \le h_0+h_1$, and the time-derivative terms $\dot{h}_0$, $j_1(z;t)$, and $j_1^*$. 

Onsager variational principle states that the time evolution equations are given by the variation of the Rayleighian with respect to the time-derivative terms. 

The variation of $\partial \mathscr{R} / \partial j_1 = 0$ leads to 
\begin{equation}
\label{variation_j1}
  f''(s) \frac{\partial s}{\partial z} + \zeta(s) j_1(s) = 0 \quad \Rightarrow \quad
  j_1(s) = -\frac{f''(s)}{\zeta(s)} \frac{\partial s}{\partial z} \, .
\end{equation}
Combining with conservation equation~(\ref{eq3}), we obtain
\begin{equation}
\label{profile_s}
  \frac{\partial s}{\partial t} = \frac{\partial }{\partial z} \Big( D(s) \frac{\partial s}{\partial z} \Big),
\end{equation}
where $D(s)$ is the diffusion constant given by
\begin{equation}
\label{D_s}
  D(s) = \frac{f''(s)}{\zeta(s)}.
\end{equation}

The boundary conditions are
\begin{equation}
\label{boundary_moving}
  s(h_0) = s^*,\quad s(h_0+h_1) = 0.
\end{equation}
Eq.~(\ref{boundary_moving}) denotes a moving boundary condition, we perform a change of variables $z' = z - h_0$, $\tau = t$.
Then Eq.~(\ref{profile_s}) becomes
\begin{equation}
\label{PDE_s}
  \frac{\partial s}{\partial \tau} = \frac{\partial }{\partial z'} \Big( D(s) \frac{\partial s}{\partial z'} \Big) + \dot{h}_{\rm 0}\frac{\partial s}{\partial z'},
\end{equation}
The boundary conditions become
\begin{equation}
\label{boundary_not_moving}
  s(z'= 0) = s^*,\quad s(z'=h_1) = 0.
\end{equation}

The variations  $\partial \mathscr{R} / \partial j_1^* = 0$ and $\partial \mathscr{R} / \partial \dot{h}_0 = 0$ lead to
\begin{align}
\label{variation_j1_star}
  & f'(s^*) + \zeta(1)h_0 \big( j_1^* + (1-s^*)\dot{h}_0 \big) = 0. \\
\label{variation_h0_dot}
  & f(1) - f(s^*) + \zeta(1)h_0 \big( j_1^* + (1-s^*)\dot{h}_0 \big)(1-s^*) = 0.
\end{align}
Combining Eqs.~(\ref{variation_j1_star}) and (\ref{variation_h0_dot}), we can derive the definition of $s^*$
\begin{equation}
\label{definition_s_star}
  \frac{f(1) - f(s^*)}{1-s^*} = f'(s^*).
\end{equation}
This is consistent with our previous definition \cite{YuTian2018}. 

Combining Eqs.~(\ref{variation_j1}), (\ref{boundary_not_moving}), (\ref{variation_h0_dot}) and (\ref{definition_s_star}), we can get the evolution equation for $h_0$
\begin{equation}
\label{ODE_h0}
  h_0 \bigg( - D(s^*)\frac{\partial s}{\partial z'} \bigg | _{z'=0} + (1-s^*) \dot{h}_0 \bigg ) = - \frac{f'(s^*)}{\zeta(1)}.
\end{equation}

To summarize, we have two coupled time-evolution equations: 
One is a PDE (\ref{PDE_s}) for the finger saturation $s(z',t)$ with the boundary conditions (\ref{boundary_not_moving}), which requires $\dot{h}_0$ for the bulk to solve. 
The second one is an ODE (\ref{ODE_h0}) for the bulk length $h_0$, which requires the finger saturation slope at the entrance ($\partial s/\partial z' |_{z'=0}$ term) to solve. 
This model is general and can be applied to any tubes of uniform cross-sections. 
The tube properties come into the model through the free energy density $f(s)$ and the friction coefficient $\zeta(s)$.

%===============================
\subsection{Bulk-only case}

If we ignore the finger flow, the Rayleighian only contains the bulk contribution
\begin{equation}
\label{eq19}
  R = f(1) \dot{h}_{\rm 0} + \frac{1}{2} \zeta(1) h_0 \dot{h}_0^2.
\end{equation}
The time evolution is given by $\delta R / \delta \dot{h}_0 = 0$, which leads to
\begin{equation}
\label{eq20}
  h_0 \dot{h}_0 = - \frac{f(1)}{\zeta(1)}.
\end{equation}
The solution to Eq.~(\ref{eq20}) with the initial condition $h_0 (t = 0) = 0$ is the Lucas-Washburn law $h_0 = C_{\mathrm{LW}} t^{1/2}$ with the prefactor 
\begin{equation}
\label{eq21}
  C_{\mathrm{LW}} = \sqrt{ \frac{2 |f(1)|}{\zeta(1)} }.
\end{equation}

%Since we have got the dynamic evolutionary equations, as long as we get needed free energy density and friction coefficients, then we can obtain the solutions. 

%=================================================
\subsection{Free energy for regular polygon tube}

\begin{figure}[htbp]
  \centering
  \includegraphics[width=0.8\columnwidth]{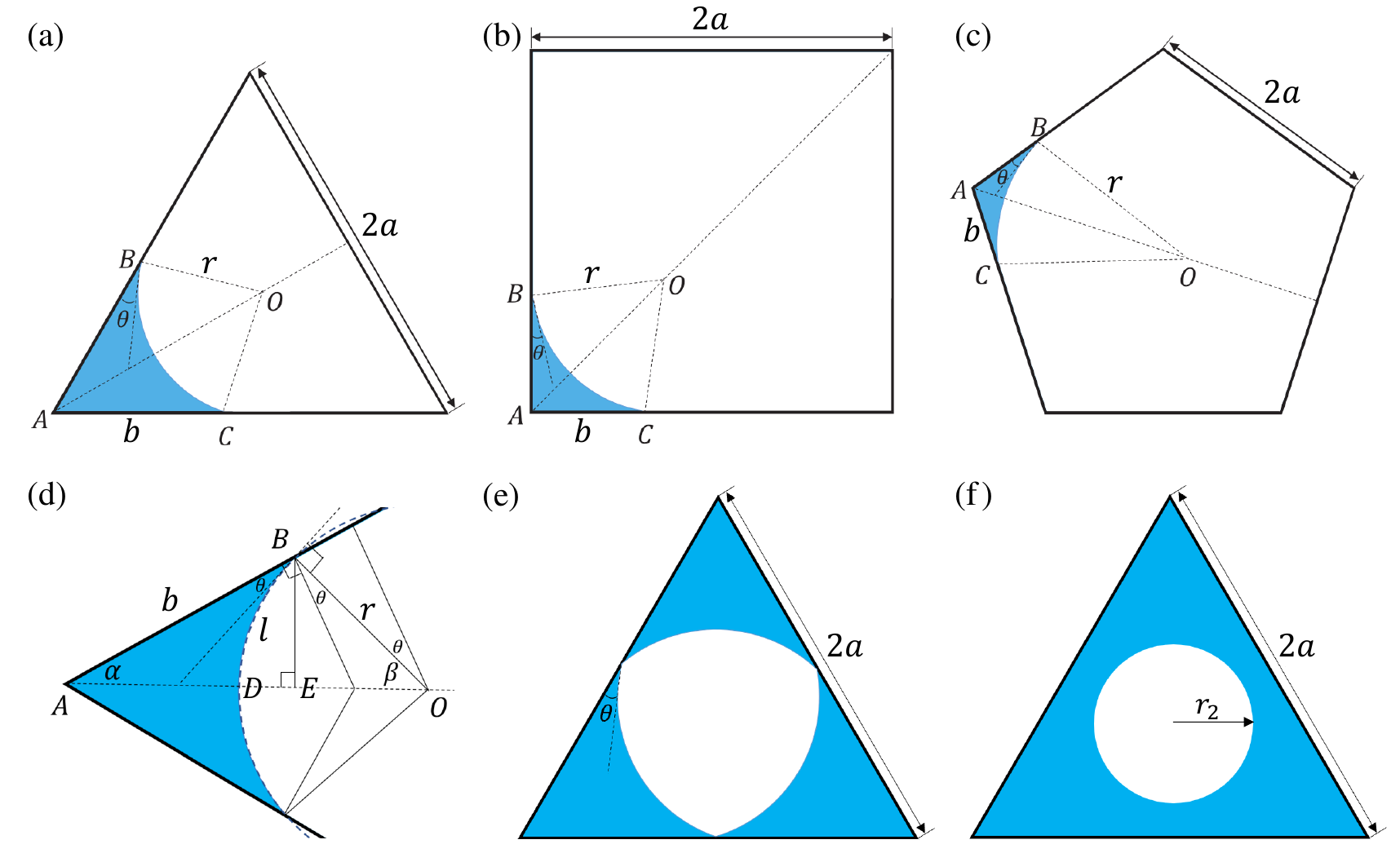}
  \caption{Sketch of the saturation of one corner in the cross-section of regular-polygon tubes. Here we take triangular, square and regular-pentagon cross-sections as examples, showing the saturation state of one corner in (a), (b) and (c). (d) shows the detailed geometrical relations in one corner. The triangles are used as examples to show two critical cases in (e) and (f). (e) expresses the critical point when the solid-vapor interface is about to disappear, which gives the critical value of the saturation as $s_{c1}$. (f) presents the liquid-vapor interface is a full circle, with the perimeter of the cross-section all covered by the fluid, and another critical value here is given by $s_{c2}$ when $r_2 = a \tan \alpha$. }
  \label{sketch3}
\end{figure}

For tubes of regular polygon cross-section, we use an integer $n$ to denote the number of edges, i.e. $n=3, 4, 5$ correspond to the triangular, square and regular-pentagon tubes, respectively.
%Now we try to derive the free energy density $f(s,n)$. The expression of $f(s,n)$ means for a given number of $n$, we can directly obtain $f(s)$.
Figure \ref{sketch3}(a)-(c) show the regular-polygon geometries, and only one corner is shown to be filled by liquid for clarity. 
%Fig \ref{sketch3}(d) shows the detailed geometrical relation.
The side length of the polygon is $2a$, and the interior angle is $2\alpha$. 
Since the sum of angles inside the regular-polygon is $(n-2)\pi$, we get the half angle as $\alpha = \pi / 2 - \pi / n$. 

%the radius of curvature of the liquid-vapor interface is denoted by $r$, the length of the curve $\overset{\LARGE{\frown}}{BD}$ is $l$, the inner angle $\angle AOB$ in the triangle $\triangle ABO$ is $\beta$, and the contact angle between the fluid and the tube wall is $\theta$. 

The finger contacts the solid surface with a contact angle $\theta$. 
The inner angle $\angle AOB$ of the triangle $\triangle ABO$ is then $\beta = \pi / n - \theta$. 
We denoted the wetting length on the polygon side $\overline{AB} = b$ (see Fig.~\ref{sketch3}(d)), then the inscribed radius of one wetted corner $r$ can be calculated as
\begin{equation}
\label{r_radius}
  r = \frac{ \overline{BE} }{ \sin \beta } = \frac{ \overline{AB}  \sin \alpha }{ \sin \beta } \nonumber 
    = \frac{ b \cos (\pi/n)}{\sin (\pi/n - \theta)} \, .
\end{equation}
Then the area occupied by the fluid for one corner is the double area of triangle $\triangle ABO$ minus the hollow sector in $\triangle ABO$
\begin{align}
\label{S_corner}
  S_\mathrm{corner} =& 2 \left( \overline{AB} \cdot \overline{BO} \cos \theta /2 - \beta r^2 /2 \right)  \nonumber \\
 =& b^2 \frac{\cos (\pi/n)}{\sin (\pi/n - \theta)} \Big( \cos \theta - (\pi/n - \theta) \frac{\cos (\pi/n)}{\sin (\pi/n - \theta)} \Big).
\end{align}
For $n$-edges regular polygon, the total area is a constant as
\begin{align}
\label{eq24}
  S_0 = n a^2 \frac{\cos (\pi / n)}{\sin (\pi / n)},
\end{align}
from which we can get the saturation $s$ as
\begin{equation}
\label{expression_s}
  s = \frac{n S_\mathrm{corner}}{S_0} 
= \left( \frac{b}{a} \right) ^2 \frac{\sin (\pi / n)}{\sin (\pi/n - \theta)} \left( \cos \theta - (\pi/n - \theta) \frac{\cos (\pi/n)}{\sin (\pi/n - \theta)} \right).
\end{equation}

The equilibrium meniscus curvature in polygonal tubes, which governs the geometric relationships derived here, aligns with classical analyses such as those by Mason and Morrow \cite{Mason1984}, who established the interplay between corner geometry and capillary equilibrium in angular cross-sections. 
Since there are solid surfaces exposed to the vapor, the range of $b$ is limited by $b < a$ [Fig. \ref{sketch3}(e)]. 
This leads to the constraint on the saturation as $0 \leq s  < s_{c1}$, and $s_{c1}$ is
\begin{align}
\label{s_c1}
  s_{c1} = s(b=a) = \frac{\sin (\pi / n)}{\sin (\pi/n - \theta)} \left( \cos \theta -  (\pi/n - \theta) \frac{\cos (\pi/n)}{\sin (\pi/n - \theta)} \right).
\end{align}

From Eq.~(\ref{expression_s}), we rewrite $b$ as a function of the saturation $s$
\begin{equation}
\label{a_s_relation}
  b = a \left[ \frac{\sin (\pi / n)}{\sin (\pi/n - \theta)} \left( \cos \theta - (\pi/n - \theta) \frac{\cos (\pi/n)}{\sin (\pi/n - \theta)} \right) \right]^{-1/2} s^{1/2}.
\end{equation}
The arc length of $\overset{\LARGE{\frown}}{BD}$ is
\begin{align}
\label{l_length}
  l = r \beta = b \frac{ \cos (\pi/n)}{\sin (\pi/n - \theta)} (\pi/n - \theta).
\end{align}

Using Eqs.~(\ref{a_s_relation}) and (\ref{l_length}), the free energy density can be expressed as a function of $s$ 
\begin{align}
\label{f_s_small_general}
  f(s) &= 2 n \left( b (\gamma _{SL} - \gamma _{SV} ) + l \gamma \right) \\
\label{f_s_small_detail}
   \frac{f(s)}{a\gamma} &= -2 n \sqrt{ \left( \frac{\sin (\pi/n - \theta)}{\sin (\pi / n)} \cos \theta - \frac{(\pi/n - \theta)}{\tan (\pi / n)} \right)} s^{1/2},
\end{align}
where $\gamma _{SL}$ and $\gamma _{SV}$ are the interfacial tensions between solid-liquid and solid-vapor, respectively. 
In this study, we assume a smooth inner wall of the tube and neglect surface roughness to simplify the analysis and isolate the geometric effects on capillary imbibition dynamics. In cases where the tube's inner wall exhibits small-scale roughness (much smaller than the tube radius), the influence of roughness can be effectively captured by adjusting the contact angle \cite{Bico2001}. Under such conditions, our model remains valid. 
In Eq. \ref{f_s_small_detail}, we have used Young's equation and chosen a dried surface as the reference point for the free energy. 
%For the case of total wetting, there might be a precursor film of molecular thickness \cite{YuTian2021,deGennes1985,Bonn2009,Tanner1979}. The first term in Eq.(\ref{f_s_small_general}) is modified as $2nb \left( \gamma _{SL} - \left( \gamma + \gamma _{SL} \right) \right)$, and the final expression remains valid. 

When the solid-vapor interfaces disappear, the liquid-vapor interface becomes a full circle of radius $r_2$ [Fig.~\ref{sketch3}(f)]. 
The saturation is given as a function of $r_2$.
\begin{equation}
\label{s_hollow}
  s = 1 - \frac{\pi r_2 ^2}{n a^2 } \tan (\pi/n ),
\end{equation}
from which we can get 
\begin{equation}
\label{r2_s_relation}
  \frac{r_2}{a} = \sqrt{ \frac{n}{\pi} \frac{1}{\tan (\pi / n)} (1 - s)}.
\end{equation}
The radius is constrained by the shortest distance from the center to the polygon side.
%which can be calculated in a universal form for regular $n-sides$ shape as $b/\cos \alpha \times \sin \alpha = b / \tan (\pi/n) $. 
This leads to a lower bound for the saturation as $s_{c2} \leq s  \leq 1$, and $s_{c2}$ is
\begin{equation}
\label{s_c2}
  s_{c2} = 1 - \frac{\pi / n}{\tan (\pi/n )}.
\end{equation}
When the contact angle $\theta = 0$, we get $s_{c1}=s_{c2}$. 

The free energy density then can be derived
\begin{align}
\label{f_s_hollow}
  f(s) &= 2 n a (\gamma _{LV} - \gamma _{SV} ) + 2 \pi r_2 \gamma  \nonumber \\
  \frac{f(s)}{a\gamma} &= -2 \left( n \cos \theta - \sqrt{ \frac{n \pi}{\tan (\pi / n)} (1 - s)} \right).
\end{align}
The free energy density at full saturation $(s = 1)$ is
\begin{equation}
\label{f_s_1}
  f(1) = -2 a \gamma \, n \cos \theta.
\end{equation}

As a summary, we show the free energy density for different cases in dimensionless form as the function of $n$, $\theta$, and $s$, which can be expressed as
\begin{equation}
\label{f_s_all}
  \frac{f(s)}{a \gamma} =
\left \{
\begin{array}{ll}
  \displaystyle -2 n \sqrt{ \left( \frac{\sin (\pi/n - \theta)}{\sin (\pi / n)} \cos \theta - \frac{(\pi/n - \theta)}{\tan (\pi / n)} \right)} s^{1/2}, \qquad \qquad & 0 \leq s \le s_{c1} \\
  \displaystyle -2 \left( n \cos \theta - \sqrt{ \frac{n \pi}{\tan (\pi / n)} (1 - s)} \right), \qquad \qquad & s_{c2} \leq s \leq 1
\end{array}
\right.
\end{equation}

%=================================================
\subsection{The condition for the existence of fingers}

\begin{figure}[htbp]
  \centering  \includegraphics[width=1.0\columnwidth]{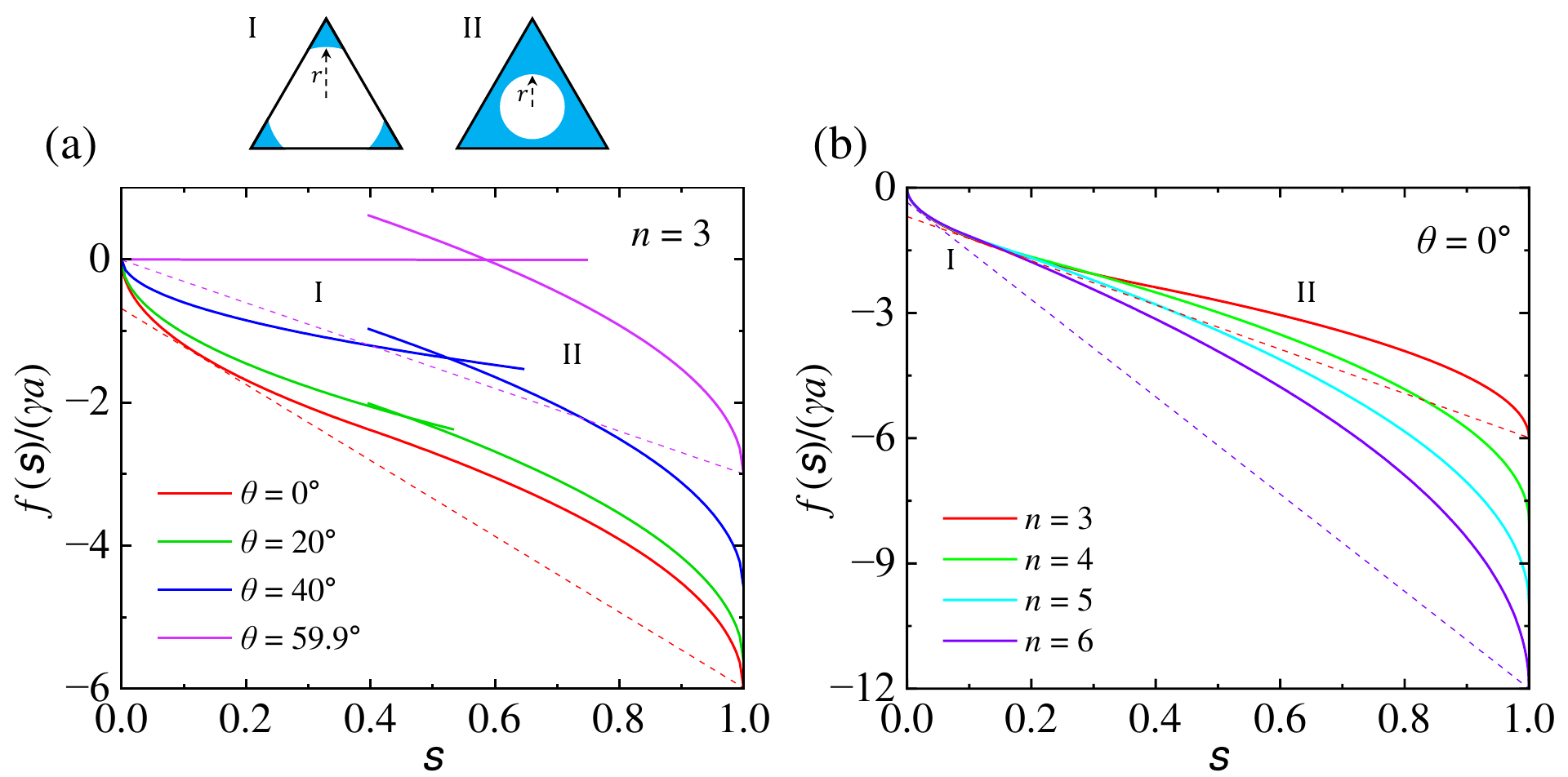}
  \caption{The free energy density curves for different $\theta$ and $n$. Two cases are shown at the top of Fig.~\ref{f_s}(a), case I is for free energy density curves at $0 \leq s \le s_{c1}$ and case II is for the curves at $s_{c2} \leq s \leq 1$. (a) Setting $n = 3$, we get different curves for $\theta = 0^{\circ}, 20^{\circ}, 40^{\circ}, 59.9^{\circ}$. The dashed lines pass through $(1,f(1))$ and tangent with the left curves at $(s^*,f(s^*))$ on $\theta = 0^{\circ}$(the red dashed line, $s^* = 0.1251$) and $\theta = 59.9^{\circ}$(the pinkish dashed line, $s^*$ is about zero).
  %The reason setting $\theta = 59.9^{\circ}$ instead of $\theta = 60^{\circ}$ is there are two $\sin (\pi/n - \theta)$ terms appear in denominators, and if $\theta = 60^{\circ}$, the value cannot be calculated.
  (b) Setting $\theta = 0^{\circ}$, we get different curves for $n = 3, n = 4, n = 5, n = 6$. The dashed grey lines pass through $(1,f(1))$ and tangent with the left curves at $(s^*,f(s^*))$ on $n = 3$(the red dashed line, $s^* = 0.1251$) and $n = 6$(the purple dashed line, $s^* = 0.0244$).}
  \label{f_s}
\end{figure}

Now we derive the condition for the existence of fingers. We plot free energy density $f(s)$ in Fig.~\ref{f_s} for different contact angle as well as different $n$. 
For the capillary imbibition system, the fully wetted state($s=1$) is a stable phase. 
We draw a straight line (the dotted line in Fig.~\ref{f_s}) goes through $(s=1,f(s=1))$ and tangential to the point at $(s=s^*,f(s=s^*))$ in the region of $s \in [0, s_{c1}]$, the value $s=s^*$ denotes a relatively stable saturation occupied by the liquid in the finger region. 
Then the stability of the capillary system comes from the coexistence of the bulk($s=1$) and the finger($s = s^*$), in which the total energy of the system keeps at the minimum state. 
The slope of the straight line is equal to the first derivative of the free energy density at the point of the equilibrium saturation $s^*$ \cite{YuTian2018, YuTian2021, ZhaoChen2021, ZhaoChen2022}, which is the same condition defined by Eq.~(\ref{definition_s_star}). 
In the following consideration, the finger starts from saturation $s^*$ at $z=h_0$ and approaches zero saturation at the finger's tips $z=h_0+h_1$.
 
Using Eq.~(\ref{f_s_all}), we obtain the free energy density of the bulk and the finger as
\begin{align}
\label{A1}
  f(s) &= -A_1 s^{1/2} a \gamma , \qquad A_1 = 2 n \sqrt{ \left( \frac{\sin (\pi/n - \theta)}{\sin (\pi / n)} \cos \theta - \frac{(\pi/n - \theta)}{\tan (\pi / n)} \right)}, \\
\label{A0}
f(1) &= -A_0 a \gamma , \qquad A_0 = 2 n \cos \theta.
\end{align}
Combined with the definition of $s^*$, we can derive it as
\begin{align}
\label{s_star_A1_A0}
  s^* = \Bigg( \frac{A_0 - \sqrt{A_0 ^2 - A_1 ^2}}{A_1} \Bigg) ^2 \, .
\end{align}
%where $s^*$ is determined by $n$ and $\theta$. 
The condition for the existence of finger corresponds to $s^*>0$, then we get $A_1>0$. 
From the expression of Eq.~(\ref{A1}), we derive the following relation
\begin{equation}
\label{verify}
  \frac{\sin (\pi/n - \theta)}{\sin (\pi / n)} \cos \theta - \frac{(\pi/n - \theta)}{\tan (\pi / n)} > 0 
  \quad \Rightarrow \quad
  \frac{\cos \theta}{\cos (\pi / n)} > \frac{(\pi/n - \theta)}{\sin (\pi/n - \theta)} \geq 1.
\end{equation}
For the function of $x$ and $\sin x$, $\frac{x}{\sin x} \geq 1$ is satisfied in the region of $x = (\pi/n - \theta) \in [-\pi, \pi]$, so at least $\cos \theta > \cos (\pi / n)$, resulting the condition for the existence of the finger as $\theta < \pi/n $. 
In Fig.~\ref{f_s}(a), we plot the free energy density curves of different contact angles and choose the maximum value as $\theta = 59.9^{\circ}$ for $n=3$. 
We see that as long as the contact angles are smaller than $\theta = 60^{\circ}$, the dashed lines passing through $\left(1,f(1) \right)$ can always tangent to free energy density curves of the finger at $\left(s^*,f(s^*) \right)$, which means the corner flows exist. 
As the value of the contact angle increases in the region $\theta \in [0^{\circ}, 60^{\circ})$, $s^*$ tends to decrease from 0.1251 to 0. 
Figure \ref{f_s}(b) also shows the free energy density curves but setting the contact angle as zero and varying $n$. 
The two dashed lines go through the two points $(s^*,f(s^*))$ and $(1,f(1))$ for $n=3$ ($s^* = 0.1251$) and $n=6$ ($s^* = 0.0244$), respectively. 
The value of $s^*$ tends to decrease as $n$ increases. 

\begin{figure}[htbp]
  \centering  \includegraphics[width=1.0\columnwidth]{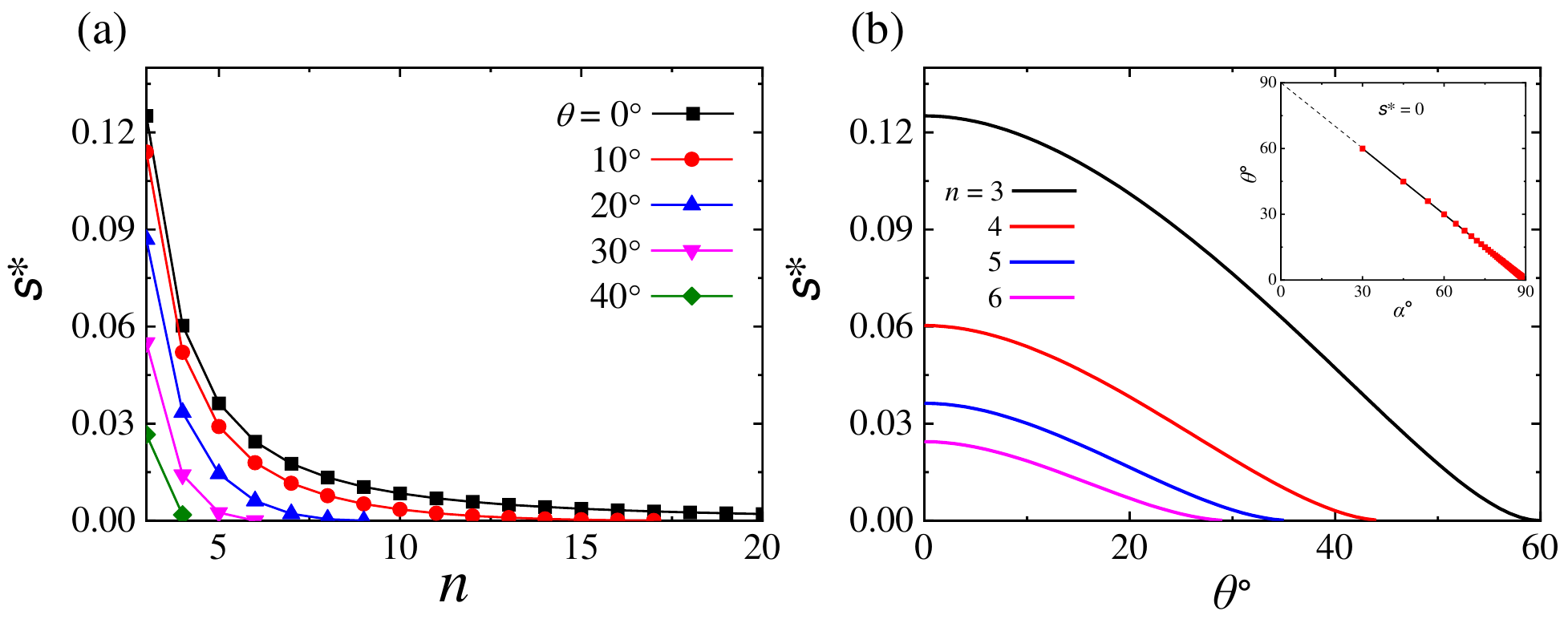}
  \caption{The relation of $s^*$ with $n$ and $\theta$. (a) The change of $s^*$ with $n$ for different contact angles as $\theta = 0^{\circ}, 10^{\circ}, 20^{\circ}, 30^{\circ}$ and $40^{\circ}$. Since $n$ is integer, we show it by a point plot, and the $x$-axis starts from 3. (b) The change of $s^*$ with $\theta$ for different $n$, Here we take $n = 3, 4, 5, 6$ as examples. The inset is the relation between $\theta$ and $n$ for $s^* = 0$, also the $x$-axis starts from 3. }
  \label{s_star_figure}
\end{figure}

Combining Eqs.~(\ref{A1}), (\ref{A0}) and (\ref{s_star_A1_A0}), we visualize the value of $s^*$ as the function of $n$ and $\theta$ in Fig.~\ref{s_star_figure} by keeping one variable unchanged. 
Figure \ref{s_star_figure}(a) fixes a set of values of contact angles as $\theta = 0^{\circ}, 10^{\circ}, 20^{\circ}, 30^{\circ}$ and $40^{\circ}$, it illustrates as $n$ increases, $s^*$ tends to decrease until to $0$ for each contact angle. 
It can be explained as $n$ becomes bigger enough, the regular polygon approaches to circular shape, while there exists no finger flow in the circular tube. 
The biggest value locates in $n=3$, which states the finger flow performs the biggest body in triangular tube compared to any others. 
As the value of the contact angle increases, $s^*$ reaches to 0 at smaller $n$.
Figure \ref{s_star_figure}(b) shows the relation of $s^*$ with varying $\theta$ for $n = 3, 4, 5, 6$. 
The value of $s^*$ decreases as the increase of $\theta$ for each $n$ and reaches to 0, which is constrained by the existence condition($\theta < \pi/n $) of the finger. 
We show the relation between $\theta$ and $\alpha$ for $s^* = 0$ in the inset of Fig.~\ref{s_star_figure}(b), the result agrees with Concus-Finn condition as $\alpha + \theta < \pi /2$ \cite{Concus1969} for finger flow. 

We list some values of $s^*$ for various $n$ and $\theta$ in Table \ref{s_star_table}.
% the data is in agreement with the theoretical conjecture of Ref. \cite{Weislogel2012}. 

\begin{table}[htbp]
\caption{Values of $s^*$. }
\centering
\begin{tabular}{|cccccc|}
  \hline
  $n$    \qquad  & \qquad   $\theta$  \qquad  & \qquad  $s^*$  \qquad  & \qquad  $n$   \qquad  & \qquad  $\theta$  \qquad  & \qquad  $s^*$ \\
  \hline
  $3$    \qquad  & \qquad   $0$   \qquad  & \qquad  $0.1251$    \qquad  & \qquad  $4$  \qquad  & \qquad  $0$   \qquad  & \qquad  $0.0603$ \\
  $4$    \qquad  & \qquad   $0$   \qquad  & \qquad  $0.0603$    \qquad  & \qquad  $4$  \qquad  & \qquad  $10$   \qquad  & \qquad  $0.0538$ \\
  $5$    \qquad  & \qquad   $0$   \qquad  & \qquad  $0.0363$    \qquad  & \qquad  $4$  \qquad  & \qquad  $20$   \qquad  & \qquad  $0.0382$ \\
  $6$    \qquad  & \qquad   $0$   \qquad  & \qquad  $0.0244$    \qquad  & \qquad  $4$  \qquad  & \qquad  $30$   \qquad  & \qquad  $0.0191$ \\
  $8$    \qquad  & \qquad   $0$   \qquad  & \qquad  $0.0133$    \qquad  & \qquad  $4$  \qquad  & \qquad  $40$   \qquad  & \qquad  $0.0031$ \\
  $10$   \qquad  & \qquad   $0$   \qquad  & \qquad  $0.0084$    \qquad  & \qquad  $4$  \qquad  & \qquad  $44$   \qquad  & \qquad  $0.000145$   \\
  $100$  \qquad  & \qquad   $0$   \qquad  & \qquad  $0.0000823$ \qquad  & \qquad  $4$  \qquad  & \qquad  $44.5$   \qquad  & \qquad  $0.0000372$   \\
  \hline
\end{tabular}
\label{s_star_table}
\end{table}

%%%%%%%%%%%%%%%%%%%%%%%%%%%%%%%%%%%%%%%%%%%%%%%%%%%%%%%%%%%%%%%%
\subsection{The dissipation function}

%For the wetting dynamics in capillary tubes, the Reynolds number is basically quite small($Re=\rho v L / \eta \textless 50 $, here we take the values of parameters as $\rho \sim 10^3 $kg/m$^3$, $L \sim 10^{-3} $m, and $ \eta \sim 20 $ mPa$\cdot$s. As for the velocity of one liquid in a capillary tube here is assumed as $v \textless 0.05 $m/s, and $v$ could be much smaller after a bit long time). 
We only consider the case of Stokes flow where the Reynolds number is small. 
In this case, the liquid flow is dominated by the laminar flow. 
We take the fluid velocity is mostly along the tube axis, $\mathbf{v} \simeq (0, 0, v_z)$ where the $x$- and $y$-components are nearly zero.
The fluid flow satisfies the Stokes equation
\begin{equation}
   \label{eq:Stokes}
   \eta \Big{(} \frac{\partial ^2 v_z}{\partial x^2} + \frac{\partial ^2 v_z}{\partial y^2} \Big{)} = \frac{\partial p}{\partial z},
\end{equation}
where $\partial p / \partial z$ is the pressure gradient in $z$ direction.

We choose a characteristic length $L$ and a characteristic velocity \cite{Ransohoff1988}
\begin{equation}
\label{bar:u}
   \bar{u} = \frac{\eta v_z}{\displaystyle \Big{(}- \frac{\partial p}{\partial z}\Big{)} L^2 } .
\end{equation}
The Stokes equation can be made into a dimensionless form
\begin{equation}
  \label{Poisson}
  \frac{\partial ^2 \bar{u}}{\partial \bar{x}^2} + \frac{\partial ^2 \bar{u}}{\partial \bar{y}^2} = -1 .
\end{equation}
The boundary conditions are $\bar{u} = 0$ at the liquid-solid interface and $ \textbf{n} \cdot \nabla \bar{u} = 0$ at the free surface, where $ \textbf{n} $ is the normal vector of the meniscus surface. 

The friction coefficient $\xi$ obeys Darcy's law $\frac{ \partial p}{\partial z} = - \xi Q$, where $Q = \int \mathrm{d} x \mathrm{d} y \, v_z$ is the volume flux. 
We can then write the friction coefficient $\xi$ as
\begin{equation}
\label{friction}
   \xi = \frac{\eta}{L^4 \int \mathrm{d} \bar{x} \mathrm{d} \bar{y} \bar{u}} .
\end{equation}
We performed numerical calculation using the finite element method in Matlab and obtained the results of $\displaystyle \int \mathrm{d} \bar{x} \mathrm{d} \bar{y} \bar{u}$ \cite{YuTian2018, ZhaoChen2021, ZhaoChen2022}. 
In this study, we aim to explore the dynamic principles of the capillary imbibition. Subsequently, our analysis is specifically focused on the perfectly wetting case, where the contact angle $\theta = 0$. 

\begin{figure}[htbp]
  \centering
  \includegraphics[width = 1.0\columnwidth]{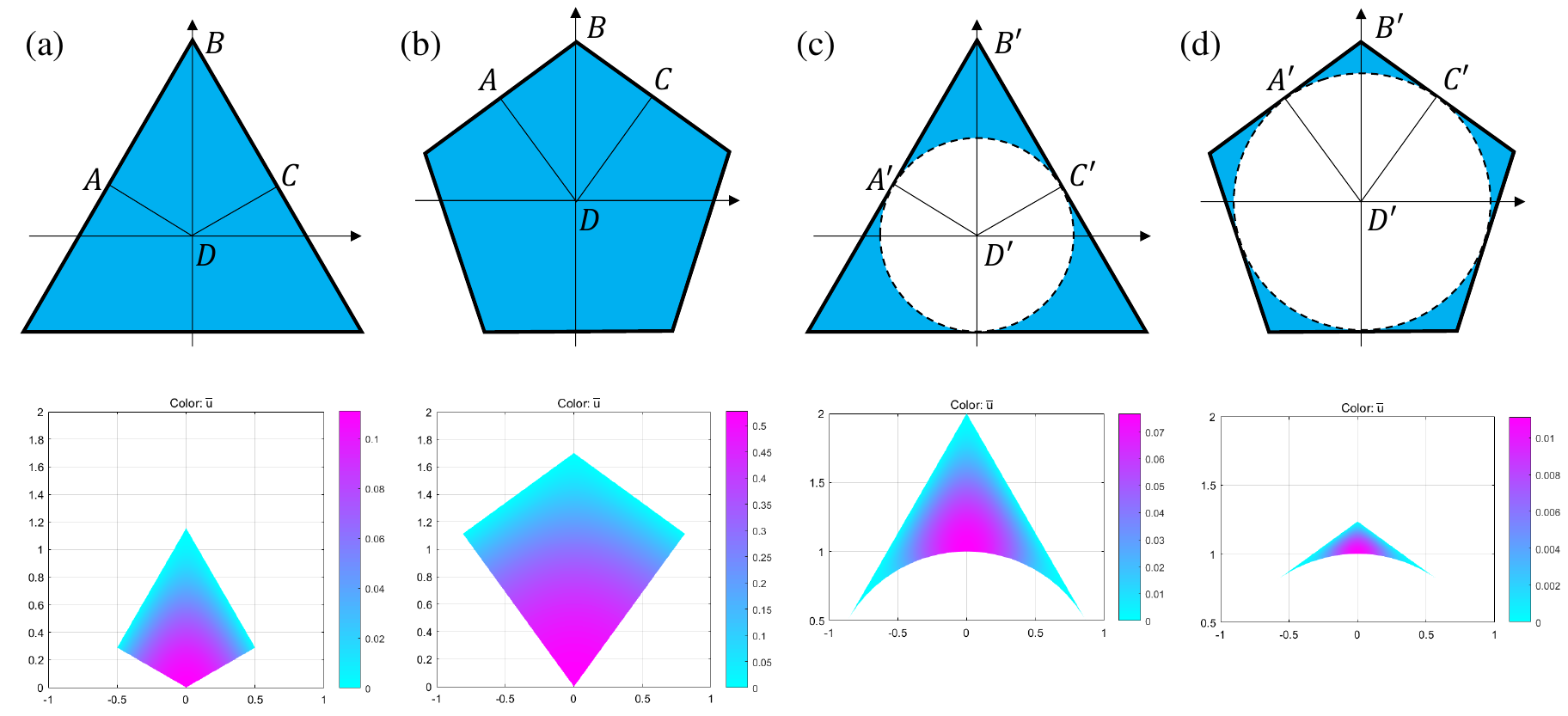}
  \caption{The sketch of geometries and Matlab results in calculating friction term, and the characteristic length is $L = a$ (here we take the regular triangle and regular pentagon as examples). (a) and (b) show the geometries and calculated results for $s = 1$, (c) and (d) show the geometries and calculated results for $s < s_{c1} $. The color map of the calculated results in the lower part shows the distribution of the dimensionless velocity component, and the scale ranges from 0 (blue) to a finite value (red), indicating the magnitude of $\bar{u}$. }
  \label{sketch_matlab}
\end{figure}

By analyzing the geometry of the regular polygon of $n$ edges, we find it's appropriate to calculate the friction of one corner region $\diamond ABCD$ (or $\diamond A'B'C'D'$) and then times $n$ to get the needed results (see Fig.~\ref{sketch_matlab}, here we take the regular triangle and regular pentagon as examples because the friction coefficient for the square case had already been calculated in our previous work \cite{YuTian2018, ZhaoChen2022}). 
The dissipation function can be expressed by the volume flux as $\displaystyle \Phi = \frac{1}{2} \int dz \xi(s)\left[ Q(z) \right]^2 $, while in the following work, we found it's more convenient to express the dissipation function in terms of the volume flux divided by the cross-section area $j(z) = Q(z)/S_0$, which is $\displaystyle \Phi = \frac{1}{2} \int dz \zeta(s)\left[ j(z) \right]^2 $. The transformation leads to a relation between the two friction coefficients by $ \zeta(s) = \xi(s) S_0 ^2$. 

For $s = 1$, we use the half-length of the side $a$ as the characteristic length. 
While for $s<s_{c1}$, we take the radius of the inscribed circle of the polygon $r$ as the characteristic length, which has been derived in Eqs.~(\ref{r_radius}) and (\ref{a_s_relation}). 
We can get the expression of coefficients for the bulk part and the finger part
\begin{align}
\label{fraction_1}
  \zeta(1) &= B_0 \eta , \qquad B_0 = \left(\frac{n}{\tan (\pi/n)} \right) ^2 \frac{1}{\int \mathrm{d} \bar{x} \mathrm{d} \bar{y} \bar{u}} , \\
\label{fraction_s}
  \zeta(s) &= B_1 s^{-2} \eta , \qquad B_1 = \left( n [\tan (\pi/n) - \pi/n ] \right) ^2 \frac{1}{\int \mathrm{d} \bar{x} \mathrm{d} \bar{y} \bar{u}} .
\end{align}
So $\zeta(1)$ and $\zeta(s)$ are determined by $n$, and Fig.~\ref{sketch_matlab} shows the calculated results of one corner by Matlab program.

\begin{table}[htbp]
\caption{Values of $B_0$ and $B_1$ for $n = 3, 4, ..., 20$. }
\centering
\begin{tabular}{|cccccc|}
  \hline
  $n$    \qquad  & \qquad   $B_0$       \qquad  & \qquad  $B_1$        \qquad  & \qquad  $n$   \qquad  & \qquad  $B_0$       \qquad  & \qquad  $B_1$ \\
  \hline
  $3$    \qquad  & \qquad   $34.6419$   \qquad  & \qquad  $61.8269$    \qquad  & \qquad  $12$  \qquad  & \qquad  $25.2713$   \qquad  & \qquad  $468.0537$ \\
  $4$    \qquad  & \qquad   $28.4544$   \qquad  & \qquad  $78.4492$    \qquad  & \qquad  $13$  \qquad  & \qquad  $25.3147$   \qquad  & \qquad  $543.3621$ \\
  $5$    \qquad  & \qquad   $26.7683$   \qquad  & \qquad  $105.5187$   \qquad  & \qquad  $14$  \qquad  & \qquad  $25.2078$   \qquad  & \qquad  $624.5281$ \\
  $6$    \qquad  & \qquad   $26.0712$   \qquad  & \qquad  $139.4177$   \qquad  & \qquad  $15$  \qquad  & \qquad  $25.3982$   \qquad  & \qquad  $711.5763$ \\
  $7$    \qquad  & \qquad   $25.7345$   \qquad  & \qquad  $179.4429$   \qquad  & \qquad  $16$  \qquad  & \qquad  $25.2337$   \qquad  & \qquad  $804.5131$ \\
  $8$    \qquad  & \qquad   $25.5518$   \qquad  & \qquad  $225.4702$   \qquad  & \qquad  $17$  \qquad  & \qquad  $24.8113$   \qquad  & \qquad  $903.3483$ \\
  $9$    \qquad  & \qquad   $25.4359$   \qquad  & \qquad  $277.3322$   \qquad  & \qquad  $18$  \qquad  & \qquad  $25.0103$   \qquad  & \qquad  $1008.0722$   \\
  $10$   \qquad  & \qquad   $25.3853$   \qquad  & \qquad  $355.0409$   \qquad  & \qquad  $19$  \qquad  & \qquad  $24.6321$   \qquad  & \qquad  $1118.6972$   \\
  $11$   \qquad  & \qquad   $25.2620$   \qquad  & \qquad  $398.6205$   \qquad  & \qquad  $20$  \qquad  & \qquad  $25.5126$   \qquad  & \qquad  $1235.2267$   \\
  \hline
\end{tabular}
\label{B0B1}
\end{table}

The detailed values of $B_0$ and $B_1$ for $n = 3, 4, ..., 20$ are showed in Table \ref{B0B1}. 
For partially saturated state, the value of friction coefficient $B_1$ is monotonically increasing. While minor numerical perturbations of $B_0$ observed for $n > 11$ arise from the discretization of high-curvature boundaries in polygonal geometries, which is inherent to finite element approximations. These perturbations do not affect the overall trend of the friction coefficient converging to the circular tube limit.

%%%%%%%%%%%%%%%%%%%%%%%%%%%%%%%%%%%%%%%%%%%%%%%%%%%%%%%%%%%%%%%%
\section{Numerical Method}

%%%%%%%%%%%%%%%%%%%%%%%%%%%%%%%%%%%%%%%%%%%%%%%%%%%%%%%%%%%%%%%%%%%%%%
\subsection{Dimensionless form of coupled PDE/ODE}

We make Eqs.~(\ref{PDE_s}) and (\ref{ODE_h0}) dimensionless.
Using Eqs.~(\ref{D_s}), (\ref{A1}) and (\ref{fraction_s}), we can get the expression of $D(s)$ as
\begin{equation}
\label{eq53}
  D(s) = \frac{A_1}{4 B_1} s^{1/2} \frac{a \gamma}{\eta} \, .
\end{equation}
The equations can be made dimensionless by the following changes of variables
\begin{align}
  \tau &= \bar{t} \frac{a \eta}{\gamma}, \\
  z' &= \bar{z}' a, \\
  D &= \bar{D} \frac{a \gamma}{\eta},
\end{align}
the PDE and ODE become 
\begin{align}
\label{PDE_dimensionless}
  & \frac{\partial s}{\partial \bar{t}} = \frac{\partial }{\partial \bar{z}'} \left( \frac{A_1}{4 B_1} s^{1/2} \frac{\partial s}{\partial \bar{z}'} \right) + \dot{\bar{h}}_{\rm 0}\frac{\partial s}{\partial \bar{z}'}, \\
  \label{ODE_dimensionless}
  & \bar{h}_0 \left( - \frac{A_1}{4 B_1} (s^*)^{1/2} \frac{\partial s}{\partial \bar{z}'} \bigg | _{\bar{z}'=0} + (1-s^*) \dot{\bar{h}}_0 \right) = \frac{A_1}{2 B_0} (s^*)^{-1/2}.
\end{align}

Combined with the boundary condition Eq.~(\ref{boundary_not_moving}), we solve Eq.~(\ref{PDE_dimensionless}) by using Euler central differential method to deal with the diffusion term and upwind differential method to deal with the convective term, respectively. 

%%%%%%%%%%%%%%%%%%%%%%%%%%%%%%%%%%%%%%%%%%%%%%%%%%%%%%%%%%%%%%%%%%%%%
\subsection{Self-similar analysis}

%As for the capillary imbibition system, we have derived one ODE in Eq.(\ref{ODE_h0}) and a PDE in Eq.(\ref{PDE_s}) dominating evolutions of $h_0$ for the bulk and $h_1$ for the finger, respectively. We wonder if $h_0$ and $h_1$ follow the same time scaling law, whether the PDE can be transformed into an ODE by scaling $z'$ using $h_0$ in $s(z',t)$ and eliminating the time term. If so, we call $s(z',t)$ admits a self-similar result, and we can obtain self-similar solutions of the system except for the numerical solutions of Eq.(\ref{ODE_h0}) and (\ref{PDE_s}). 
The coupled PDE/ODE system also can be cast into a self-similar form. 
We assume $h_0$ and $h_1$ obey the time-dependence as 
\begin{equation}
\label{h0_and_h1_assume}
  h_0 (t) = {C}_\mathrm{b} t^{m}, \qquad h_1 (t) = {C}_\mathrm{f} t^{m},
\end{equation}
and the finger profile is self-similar as
\begin{equation}
\label{self_s_F}
  s (z',t) = F \left( \frac{z'}{h_0} \right) = F(\chi), \quad \chi = \frac{z'}{h_0} \, .
\end{equation}
Then the PDE (\ref{PDE_s}) becomes
\begin{equation}
\label{self_s}
  - \chi m C_\mathrm{b} \frac{\mathrm{d} F(\chi)}{\mathrm{d} \chi} t^{m-1} = \frac{1}{C_\mathrm{b}} \frac{\mathrm{d}}{\mathrm{d} \chi} \left( D \frac{\mathrm{d} F(\chi)}{\mathrm{d} \chi} \right) t^{-m} + m C_\mathrm{b} \frac{\mathrm{d} F(\chi)}{\mathrm{d} \chi} t^{m-1} .
\end{equation}
The above equation becomes time-independent only if the condition $m - 1 = - m$ is satisfied, then we can get $m = \frac{1}{2}$. 
So we derive an ODE for the finger profile $F(\chi)$ as
\begin{equation}
\label{ODE_F}
   (1 + \chi) \frac{\mathrm{d} F(\chi)}{\mathrm{d} \chi} + \frac{2}{C_\mathrm{b} ^2} \frac{\mathrm{d}}{\mathrm{d} \chi} \left( D \frac{\mathrm{d} F(\chi)}{\mathrm{d} \chi} \right) = 0 .
\end{equation}
We make the equation dimensionless by
\begin{equation}
\label{self_dimensionless_parameters}
  C_\mathrm{b} = \tilde{ \mathscr{C}_\mathrm{b}} {C}_\mathrm{LW}, \qquad D = \tilde{ D} {C}_\mathrm{LW} ^2,
\end{equation}
then Eq.~(\ref{ODE_F}) becomes
\begin{equation}
\label{ODE_F_dimensionless}
  (1 + \chi) \frac{\mathrm{d} F(\chi)}{\mathrm{d} \chi} + \frac{2}{\tilde{\mathscr{C}_\mathrm{b} ^2}} \frac{\mathrm{d}}{\mathrm{d} \chi} \left( \tilde{D} \frac{\mathrm{d} F(\chi)}{\mathrm{d} \chi} \right) = 0.
\end{equation}

The ODE (\ref{ODE_h0}) for the bulk flow becomes
\begin{align}
  - \tilde{D} \frac{\mathrm{d} F(\chi)}{\mathrm{d} \chi} \Big| _{\chi = 0} +\frac{1}{2} (1-s^*) \tilde{\mathscr{C}_\mathrm{b} ^2} = - \frac{1}{2} \frac{f'(s^*)}{|f(1)|} \nonumber \\
  \label{self_h0}
 \Rightarrow \tilde{\mathscr{C}_\mathrm{b} ^2} = \frac{1}{1-s^*} \left( 2 \tilde{D} \frac{\mathrm{d} F(\chi)}{\mathrm{d} \chi} \Big| _{\chi = 0} - \frac{f'(s^*)}{|f(1)|} \right).
\end{align}

The self-similar form has reduced the complexity of the equations. 
The PDE (\ref{PDE_s}) for the finger profile has become an ODE (\ref{ODE_F}), while the ODE (\ref{ODE_h0}) for the bulk becomes an algebraic equation (\ref{self_h0}).  

For regular-polygon tubes, the following quantities are used to make the equations dimensionless
\begin{align}
\label{C_LW-square}
  {C}_\mathrm{LW} ^2 &= \frac{2 |f(1)|}{\zeta(1)} = \frac{2 A_0}{B_0} \frac{a \gamma}{\eta}, \\
  \tilde{ D}(s) &= \frac{D(s)}{{C}_{LW} ^2} = \frac{A_1 B_0}{8 A_0 B_1} s^{1/2} = \tilde{D}_0 s^{1/2}, \qquad  \tilde{D}_0 = \frac{A_1 B_0}{8 A_0 B_1}  \\
  - \frac{f'(s^*)}{|f(1)|} &= \frac{A_1}{2 A_0} (s^*)^{-1/2}.
\end{align}
Note that the Lucas-Wahburn coefficient ${C}_\mathrm{LW}$ in Eq. (\ref{C_LW-square}) depends on $n$. Then Eqs.~(\ref{ODE_F_dimensionless}) and (\ref{self_h0}) become
\begin{align}
\label{self_F_dimensionless}
  0 &=\tilde{D}_0 \frac{\mathrm{d}}{\mathrm{d} \chi} \left( F^{1/2} \frac{\mathrm{d} F(\chi)}{\mathrm{d} \chi} \right) + \frac{\tilde{\mathscr{C}_\mathrm{b} ^2}}{2} (1 + \chi) \frac{\mathrm{d} F(\chi)}{\mathrm{d} \chi},  \\
  \label{self_h0_dimensionless}
  \tilde{\mathscr{C}_\mathrm{b} ^2} &= \frac{1}{1-s^*} \left( 2 \tilde{D}_0 (s^*)^{1/2} \frac{\mathrm{d} F(\chi)}{\mathrm{d} \chi} \Big| _{\chi = 0} + \frac{A_1}{2 A_0} (s^*)^{-1/2} \right).
\end{align}

We sketch the procedures to solve Eqs.~(\ref{self_F_dimensionless}) and (\ref{self_h0_dimensionless}): 
\begin{enumerate}
\item We first take an initial guess value for $  \mathrm{d} F(\chi) / \mathrm{d} \chi | _{\chi = 0}  = k$. 
Since $F(\chi) = s (z',t)$ is a monotonically decreasing function, so the guess value should be negative; 
\item Taking the guess value $k$ into Eq.~(\ref{self_h0_dimensionless}), we obtain the result of $\tilde{\mathscr{C}_\mathrm{b}}$;
\item Once we have $k$ and $\tilde{\mathscr{C}_\mathrm{b}}$, we can solve Eq.~(\ref{self_F_dimensionless}) and get the result of $F(\chi)$ function, which should satisfy the condition: 
\begin{align}
\label{self-condition}
  F(\chi \rightarrow \infty) \rightarrow 0. 
\end{align}
In the actual calculation, we choose one judge value $\epsilon = 10 ^{-5}$ and as long as $F(\chi \rightarrow \infty) < \epsilon$, we get the right solution. 
\item If the condition Eq.~(\ref{self-condition}) is not satisfied, then we need to adjust the guess value $ k $ and redo $step~1-3$ until the condition is satisfied;
\item The value of $\chi_1$ at the first point for $F(\chi _1) = 0$ gives the finger speed $\tilde{\mathscr{C}_\mathrm{f}} / \tilde{\mathscr{C}_\mathrm{b}} = \chi_1$. 
\end{enumerate}

%We shall show the detailed form of the self-similar solutions in the next section and compare it with the numerical results of the PDE and the ODE. 

%%%%%%%%%%%%%%%%%%%%%%%%%%%%%%%%%%%%%%%%%%%%%%%%%%%%%%%%%%%%%%%%%%%%%%%%
\section{Results and discussion}

We present the numerical solutions by solving the coupled PDE and ODE, and also achieve the self-similar solutions to give the dynamic evolutionary results of the capillary filling system, namely $h_0$ and $h_1$. All derivations are performed by the dimensionless form. 

\subsection{The numerical and self-similar results}
\begin{figure}[htbp]
  \centering
  \includegraphics[width=0.8\columnwidth]{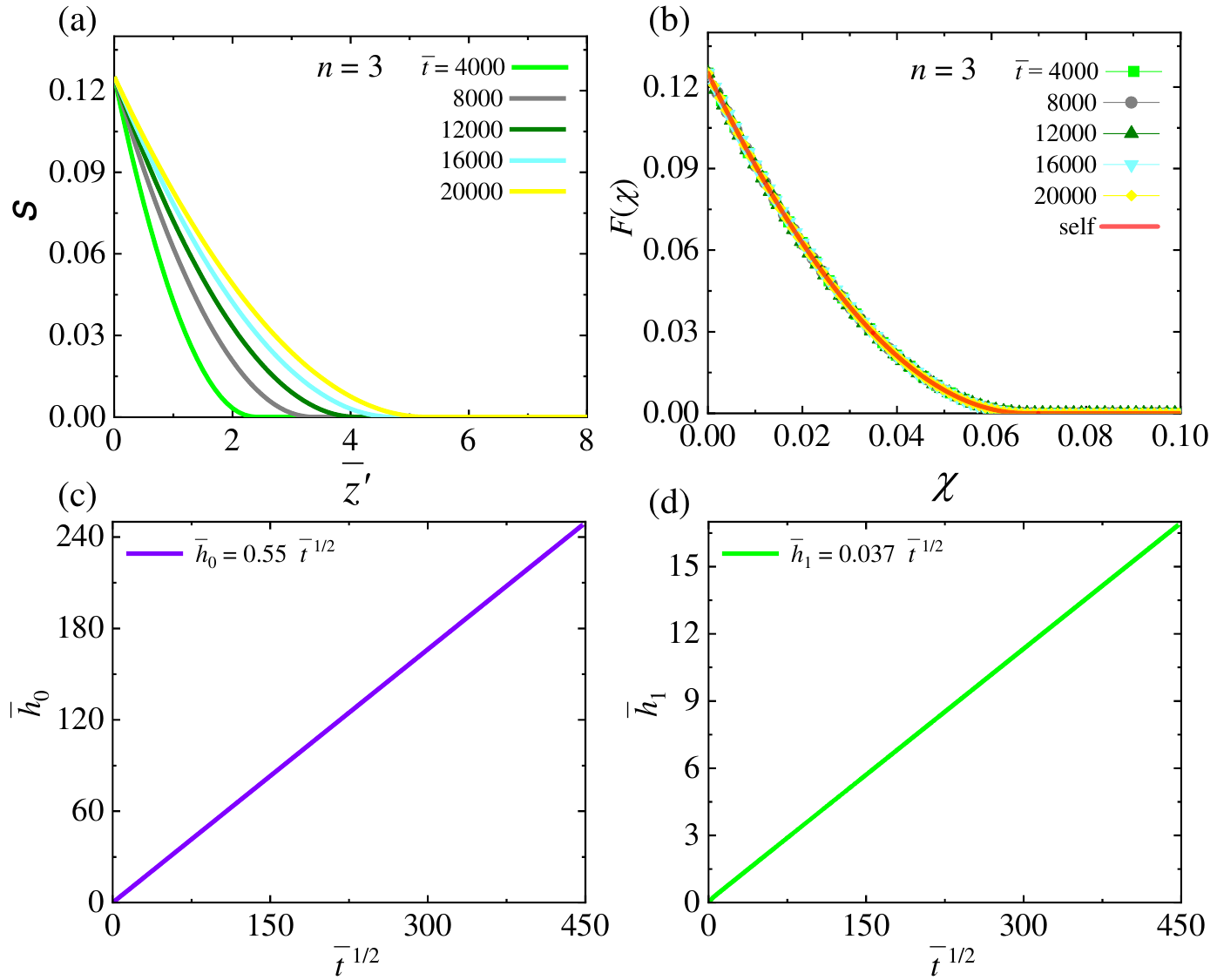}
  \caption{ Solutions for $n = 3$. (a), (c) and (d) are from numerical calculations, while (b) is the self-similar solution combined with the numerical calculation of finger profiles. (a) The saturation profiles of the finger flow at different times. (b) The self-similar solution and the saturation profiles but scaled by $h_0$ for different times. The finger profiles at different time are shown as symbols of different colors, and they all coincide by the self-similar curve. (c) The length of the bulk as a function of time, which is $\bar{h}_0 = 0.55 \bar{t}^{1/2}$. (d) The length of the finger as a function of time, which is $\bar{h}_1 = 0.037 \bar{t}^{1/2}$. }
  \label{n3}
\end{figure}

We achieve numerical solutions by solving Eqs.~(\ref{PDE_dimensionless}) and (\ref{ODE_dimensionless}), and also obtain self-similar solutions of Eqs.~(\ref{self_F_dimensionless}) and (\ref{self_h0_dimensionless}). We only take $n = 3$ as one example and show the results in Fig. \ref{n3}. 

The finger profiles at different times are shown in Fig.~\ref{n3}(a) and the result of the self-similar solution is shown in Fig.~\ref{n3}(b). Fig.~\ref{n3}(b) also indicates the saturation profiles of different times overlap to one curve and coincide with $F(\chi)$. 
The lengths of the bulk $\bar{h}_0$ and finger $\bar{h}_1$ as functions of time are shown in Fig.~\ref{n3}(c) and (d), of which both lengths follow $\bar{t}^{1/2}$ scaling law. The $\bar{t}^{1/2}$ scaling law observed for the finger liquid is consistent with the experimental findings reported in Dong's work \cite{Dong1995}, where the bulk dynamics were not considered. The evolutionary coefficients of $h_0$ and $h_1$ are 0.55 and 0.037, indicating the filling velocity of the bulk is much faster than the finger's. 

\begin{figure}[htbp]
  \centering
  \includegraphics[width = 1.0\columnwidth]{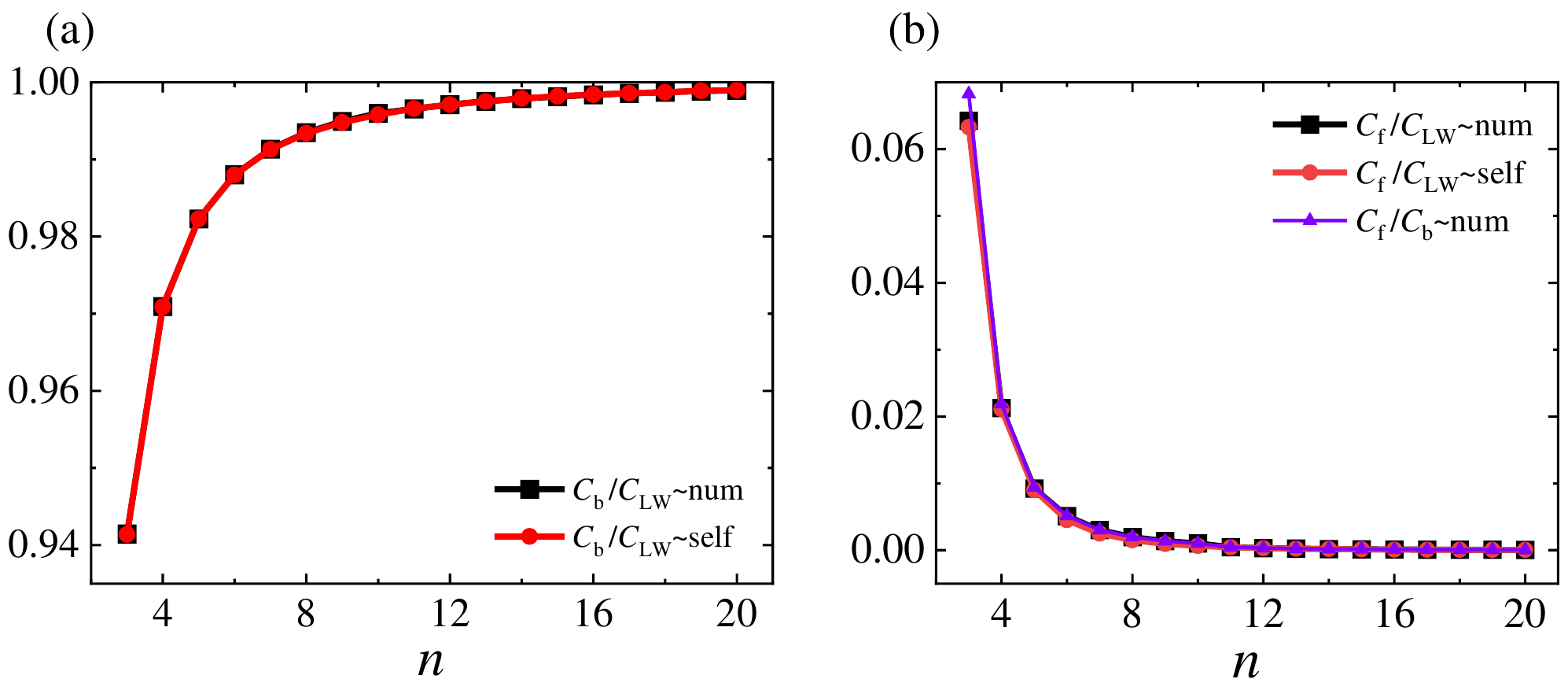}
  \caption{The evolution coefficients of the bulk flow and the finger flow change with $n$. The black line with square points represents the numerical results by solving the coupled ODE and PDE equations, the red line with dot points represents the self-similar solutions, and they are scaled by the Lucas-Washburn prefactor $C_\mathrm{LW}$. The purple line with triangular dots in Fig.~\ref{CbCf}(b) represent $C_\mathrm{f} / C_\mathrm{b}$ of numerical solutions. }
  \label{CbCf}
\end{figure}

The numerical solutions and the self-similar solutions of the evolutionary coefficients for both the bulk and the finger at $n = 3 \sim 10$ are shown in Table \ref{self_num}. 
Our results of $C_\mathrm{f} / C_\mathrm{b}$ are generally in good agreement with Weislogel's \cite{Weislogel2012}. 
We also visualize the results scaled by the Lucas-Washburn prefactor in Fig.~\ref{CbCf}, taking $n$ increases from $3$ to $20$ as $x$-axis. 
The reduction in the bulk flow velocity caused by the finger effect becomes more pronounced as the polygon side number $n$ decreases. This trend arises because smaller $n$ corresponds to larger critical finger saturation $s^*$, which intensifies the viscous dissipation in the finger region and thereby exerts stronger retardation on the bulk flow. This theoretical prediction aligns with experimental observations in open rectangular microchannels by Kolliopoulos et al. \cite{Kolliopoulos2021}. Their study explicitly demonstrated that the coupling between finger flow and bulk flow reduces the bulk velocity, primarily due to energy redistribution caused by the  pre-wetting of fingers along channel corners. 

\begin{table}[htbp]
\centering
\caption{The numerical solutions and self-similar solutions for $n = 3 \sim 10$. $C_\mathrm{b}$ and $C_\mathrm{f}$ are dimensionless filling coefficients of the bulk flow and the finger flow, which are obtained by numerically solving PDE and ODE. $\tilde{C}_\mathrm{b}$ and $\tilde{C}_\mathrm{f}$ are parameters of $C_\mathrm{b}$ and $C_\mathrm{f}$ scaled by the Lucas-Washburn prefactor. $\tilde{\mathscr{C}_\mathrm{b}}$ and $\tilde{\mathscr{C}_\mathrm{f}}$ are dimensionless filling coefficients of the bulk and the finger also scaled by the Lucas-Washburn prefactor, which are obtained from self-similar solutions. The results of Weislogel \cite{Weislogel2012} are also shown for comparison.}
\begin{tabular}{|c|c|cccccc|cc|}
  \hline
  $n$ & $C_\mathrm{LW}$ & $C_\mathrm{b}$ & $C_\mathrm{f}$ & $\tilde{C}_\mathrm{b}$ & $\tilde{C}_\mathrm{f}$ & $C_\mathrm{f} / C_\mathrm{b}$ & Weislogel & $\tilde{\mathscr{C}_\mathrm{b}}$ & $\tilde{\mathscr{C}_\mathrm{f}}$ \\
  \hline
  $3$ & $0.5886$ & $0.5541$ & $0.03729$  & $0.9414$ & $0.0634$ & $0.06735$ & $0.06925$ & $0.9414$ & $0.0633$  \\
  $4$ & $0.7499$  & $0.7278$ & $0.01574$  & $0.9705$ & $0.0210$ & $0.02164$ & $0.02959$ & $0.9709$ & $0.0211$  \\
  $5$ & $0.8644$  & $0.8486$ & $0.00780$  & $0.9817$ & $0.0090$ & $0.00917$ & $0.01369$ & $0.9823$ & $0.0090$  \\
  $6$ & $0.9595$  & $0.9480$ & $0.00489$  & $0.9880$ & $0.0051$ & $0.00516$ & $0.00702$ & $0.9880$ & $0.0045$  \\
  $7$ & $1.0431$  & $1.0341$ & $0.00315$  & $0.9913$ & $0.0030$ & $0.00302$ & - & $0.9913$ & $0.0025$  \\
  $8$ & $1.1191$  & $1.1118$ & $0.00220$  & $0.9935$ & $0.0020$ & $0.00201$ & $0.00235$ & $0.9934$ & $0.0015$  \\
  $9$ & $1.1897$  & $1.1837$ & $0.00163$  & $0.9949$ & $0.00137$ & $0.00137$ & - & $0.9948$ & $0.00095$  \\
  $10$ & $1.2553$  & $1.2502$ & $0.00128$  & $0.9960$ & $0.00102$ & $0.00102$ & $0.00099$ & $0.9958$ & $0.00064$  \\
  \hline
\end{tabular}
\label{self_num}
\end{table}

The numerical and the self-similar solutions are in good agreement with each other in Fig.~\ref{CbCf}. When $n=3$, the discrepancy is obvious as $\tilde{\mathscr{C}_\mathrm{b}} = 0.94$ and $\tilde{\mathscr{C}_\mathrm{f}} = 0.06$, indicating the evolutionary coefficient of the bulk is reduced by the finger effect by $6 \%$. 
We can conclude when $n$ is a small value, there is a relatively big difference between the coupled result and the Lucas-Washburn coefficient, and as $n$ increases, the evolution coefficient $C_\mathrm{b}$ gradually approaches the Lucas-Washburn prefactor. 
The reason is that when $n$ is bigger enough, the regular-polygon tube corresponds to the circular tube, while the circular tube appears only the bulk flow, resulting the same velocity with Lucas-Washburn's. Fig.~\ref{CbCf}(b) shows that as $n$ increases, evolution coefficients of the finger flow decrease to $0$, which also agrees well with the capillary filling situation of the circular tube case.

%%%%%%%%%%%%%%%%%%%%%%%%%%%%%%%%%%%%%%%%%%%%%%%%%%%%%%%%%%%%%%%%%%%%%%%%%%%%%%
\section{Conclusion}

We have studied the capillary dynamics of the wetting liquid in regular-polygon tubes. 
Using Onsager variational principle, we derive an ODE for the bulk flow and a PDE for the finger profile, and these two equations couple with each other. 
The coupled ODE/PDE has a self-similar solution, and we obtain both the numerical solution to coupled time evolutionary equations directly and also the self-similar solution
The two results agree with each other and show the evolutionary length of the bulk and the finger follows a $t^{1/2}$ scaling law.
%The good coincidence of the two solutions can mutually verify the correctness, which is of great help to deal with corresponding problems. 

We scale the filling velocity by the Lucas-Washburn prefactor, and the results manifest the existence of the finger slows down the filling speed of the bulk under the condition of the coupling effect. 
The reduction is characterized by a starting finger saturation $s^*$. 
Larger $s^*$ causes the slower velocity in the bulk, and the maximum value of $s^*$ presents in the triangular tube as $n=3$. 
As the number of edges of the regular-polygon $n$ increases, the filling speed gradually approaches the Lucas-Washburn prefactor.
When the value of $n$ is large enough, the evolutionary situation in the regular-polygon tube is essentially the same with in the circular tube, where the fingers just disappear.

\begin{acknowledgments}
We acknowledge the support of the National Key R\&D Program of China (2022YFE0103800), the National Natural Science Foundation of China (22373036), and SRP project of SCUT (X202310561380).  
The computation was made possible by the facilities of Information and Network Engineering and Research Center of SCUT.
\end{acknowledgments}

\bibliography{wetting}

\end{document}